Full paper

# Effective Reduction of Oxygen Debris in Graphene Oxide

Orit Seri-Livni[1,2], Cecile Saguy[2], Faris Horani[1,2], Efrat Lifshitz[1,2#], Dima Cheskis*

[#] O. Seri-Livni, Dr. C. Saguy, F. Horani, Prof. E. Lifshitz

[1] Schulich Faculty of Chemistry

[2] Solid State Institute

Russell Berrie Nanotechnology Institute

Technion, Haifa 3200003, Israel

E-mail: ssefrat@technion.ac.il

[*] Dr. D. Cheskis

Physics Department

Ariel University, Ariel 407000, Israel

E-mail: dimach@ariel.ac.il

## Abstract

Graphene oxide (GO) raised substantial interest in the last two decades thanks to its unique properties beyond those of pristine graphene, including electronic energy band-gap, hydrophilic behavior and numerous anchoring sites required for functionalization. In addition, GO was found to be a cheap mass-production source for the formation of the pristine graphene. However, the presence of numerous clusters containing oxygen functional groups (called debris) on the GO surface hinders the GO integration in electronic devices.

Here, we present a simple method aimed to reduce the density of oxygen debris weakly bonded to the surface. The method consists of minimal treatments, like sonication and/or water rinsing processes. Whereas this simple method removed epoxy and hydroxyl oxygen groups weakly attached to the graphene matrix, the double C=O bonds are almost not affected by the applied treatment, as demonstrated by X-ray photoelectron spectroscopy and Fourier transform infrared spectroscopy. Scanning tunneling microscopy and high-resolution transmission electron microscopy measures designated non-uniform distribution of the oxidation sites, appearing as clusters concentrated preferentially on GO defected regions, albeit separated by pristine graphene areas. The results should have an impact in the implementation of GO in electronic devices deposited on different substrates.

## 1. Introduction

The unique properties of graphene, a single layer of graphite, were first described theoretically by Wallace in 1947.[1] More than fifty years later, in 2004, Geim and Novoselov achieved mechanically exfoliated graphene.[2] Since then, additional methods to grow graphene were developed, such as chemical vapor deposition,[3] epitaxial growth,[4] electrochemical growth,[5] electrochemical[6] and chemical[7] exfoliations. Brodie et al. were the first to oxidize graphite, as early as 1859.[8] More than a century later, in 1962, Boehm et al. achieved graphene oxide (GO). [9] Their material included both single-layers and multilayer flakes. Since 1962, the processes of synthesis and sonication have constantly improved. Today most of the GO are produced by the modified Hummers method where graphite is first oxidized, then sonicated and lastly the amount of oxygen is decreased. During the exfoliation process, atoms of oxygen intercalate between the graphene layers, weakening the interlayer van der Waals forces. Then, a sonication process separates the exfoliated oxidized graphite into single oxidized graphene layers, called graphene oxide (GO). These layers consist of a graphene sheets bonded with oxygen functional groups.[10] As the percentage of oxygen groups on the surface changes, the electronic behavior of a GO film can change from insulator via semiconductor towards graphene with zero energy bandgap. The semiconductor electronic behavior is characterized by the parabolic dependence of the electron energy on its momentum, while the graphene electronic behavior displays a linear dependence, producing a "v-shaped" electronic energy spectrum.[2] When most oxygen groups are removed from the surface, the resulting material is called reduced graphene oxide (RGO) .[11] This RGO material, also named chemical exfoliated graphene, is the last step of the exfoliation process. RGO structurally resemble a pristine graphene, however differs from it by the residual number of oxygen atoms, with an electronic band-gap determined by the oxygen concentrations. In addition, the GO and RGO are accessible to functionality. The cheapest form for producing GO is as flakes inside water or an organic solution. It is hydrophilic, and can be deposited on various substrates. GO is used in several biomedical applications,[12-14] as catalysts,[15,16] water filters,[17] solar cells component,[18] part of lithium-ion battery device (as RGO) [19] and in supercapacitors.[20]

However, alongside all these advantages, the major disadvantage in the production of GO and RGO films is that it is difficult to control the distribution of functional groups over the surface. Many efforts have been made to understand and control the structure of deposited graphene oxide.[21] The distribution of functional groups was quantified by the Lerf–Klinowski (LK) theoretical model,[22] which is based on thermodynamic calculations. This model shows that the functional group distribution can affect many parameters, such as conductivity, energy band gap, mechanical strength and thermal effects. In this model, epoxy and hydroxyl groups are distributed randomly over the GO flake, while carboxyl sits on the corners of the flake.

There have been many experimental attempts to see the microscopic structure and microscopic parameters of GO using atomic force microscopy (AFM),[23-25] high-resolution transmission electron microscopy (HR-TEM),[26-29] scanning transmission electron microscopy (STEM),[30,31] and scanning tunneling microscopy (STM).[32-37] Scanning tunneling spectroscopy (STS) measurements were also used to study the local electronic density of states. Wang et al. showed similarity between the electronic spectrum of RGO and that of pure graphene, as deduced from STS dI/dV curves ,[37] thus, suggesting that the residual oxygen groups have limited influence on the final spectrum. In contrast, Harthcook et al. showed that in locally oxidized CVD graphene samples, it is possible to detect different electronic behavior between pure graphene (dI/dV has a "V-shaped") and the proximity of oxygen atoms (dI/dV has a parabolic shape) by local STS probing.[33] Due to the fact that oxygen groups are not arranged in any defined order, and that their distribution depends on synthesis and deposition procedures, the results of these experiments vary widely. For example, Pandey et al., showed by using STM that in RGO sample,[35] oxygen groups were aligned in rows. Katano et al. demonstrated that it is possible to perform STM measurements of graphene oxide,[34] but did not image the sample at atomic resolution due to the high density of oxygen groups. Moreover, during the synthesis and sonication processes required for producing GO, oxygen groups organized in clusters called oxygen debris (OD). The nature of the clusters and how they are bonded to the graphene oxide are still under debate. Thomas et al. developed a model showing that these OD actually changes the GO material into a system with two components: Some regions of the film are pure graphene, while other regions are covered by functional oxygen groups.[38] The OD can be washed away by a strong base such as NaOH. Controversially, Dimiev et al. said that these clusters were actually created during the strong base wash.[39] The discussion over oxygen debris (OD) has continued ever since, sustained by many authors using different techniques [40-44]. Thomas et al. [40] and Naumov et al. [41] both performed GO photoluminescence experiments, supporting and opposing the two-component model, respectively. Bonanni et al.,[42] demonstrated the influence of OD on inherent electroactivity. These researchers claimed that OD consists of highly oxidized polyaromatic fragments adsorbed to the graphene matrix by non-covalent bonds such as π-π stacking, hydrogen bonding, and van der Waals interactions. They also showed that OD can be removed by extensive sonication. Chen and Chen [45] showed that aromatic compounds have tendency to be organized by π-π interaction on defect sites of a graphene matrix.

An additional question exists about the type of bonding to a substrate. This can drastically change the electronic behavior of deposited GO films, and possibly their morphology as well. A chemical bonding to the substrate would generate a strong local distortion of the hexagonal matrix [46,47]. According to recent experiments, covalent bonds are expected to cause local changes in topography and also in the STS curves [48,49]. Strong interaction can be also seen in a Moiré pattern. For example, in an experiment where gold islands are covered by graphene, Palinkas et al., [50] observed Moiré superlattices and STS curves with different periodicities.

In this work we investigate the structural, chemical and electronic properties of the GO surface thin films and their bonding to oxygen functional groups as a function of minimal preparation treatments, like sonication and/or water rinsing processes. We show, by atomically resolved STM and HR-TEM, that OD concentrated preferentially on GO defected regions. Sonication of the GO flake dispersed in ethanol followed by simple water rinsing lead to removal of the non-covalently adsorbed oxygen debris and to the recovery of part of the C-C bonds, as shown by X-ray photoelectron spectroscopy (XPS) and Fourier transform infrared spectroscopy (FTIR). Moreover, comparison between the STM and HR-TEM images bestows that the morphology of GO after the described treatments, resembles that of suspended GO without an influence of a substrate. This is possible by assuming that part of the oxygen functional group clusters lay below the graphene film and that OD attached to the lower surface. The lack of a Moiré pattern supports this hypothesis.

## 2. Results

### 2.1 Sample preparation

Standard industrial GO flakes dispersed in an ethanol solution and prepared by a modified Hummers method were purchased. The flakes were deposited on a pre-cleaned Au/Mica substrate using a drop-cast method (see experiment section). The investigation followed the GO nature thru three different stages: (i) the as-purchased GO, (ii) after 15 min of sonication and (iii) after 15 min sonication followed by water rinsing.

### 2.2 Characterization

(i) The as-purchased GO flakes

Drop-casted GO flakes from the purchased solution without a dilution, sonication or water rinsing, which were deposited onto a Au/Mika substrate, are shown in **Figure 1**(a). AFM measurements have shown flakes up to 10 microns in width with a thickness between 5 to 14 nm, as depicts in **Figure 1**(b) in correspondence to the color bars labeled in panel (a). The observed thickness suggests agglomeration of a few layers, when a thickness of single layer is ~ 1 nm.

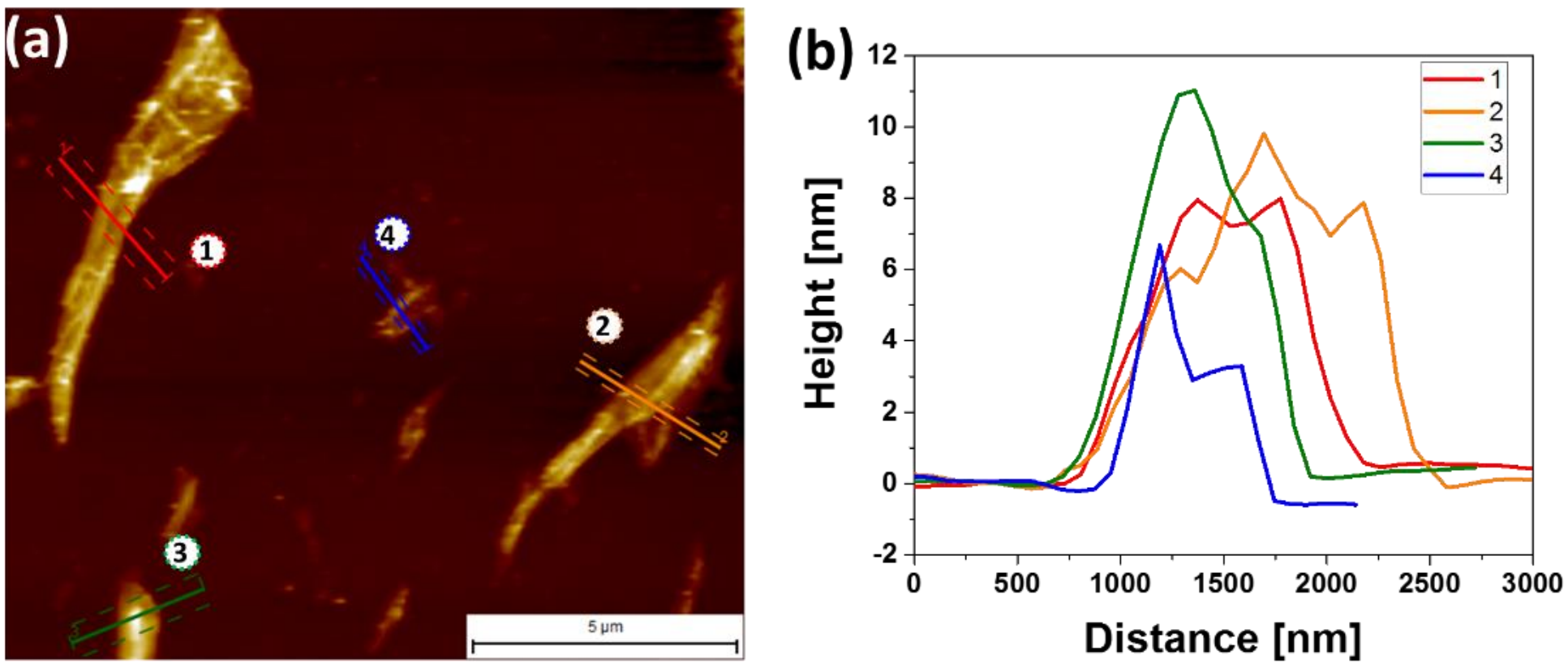

*Figure 1: Atomic Force Microscopy (AFM) of as-purchased GO flakes: (a) AFM image; (b) Height profile of the GO domains. 1, 2, 3 and 4 are line profiles of random GO domains, synchronized with the marked flakes in panel (a).*

Raman spectroscopy measurements of the as-purchased GO material are given at the supplement information (SI, Figure S1), exposing the expected resonance transitions that reflects the GO composition. [51]

The STM image (Figure 2 (a)) shows bright patterns superimposed onto the graphene atomically resolved structure. The randomly dispersed bright spots are most probably related to agglomeration of oxygen functional groups (e.g., epoxy, carboxyl and also hydroxyl). Indeed, the fast Fourier transform (FFT) (Figure 2 (b)) of the STM image exhibits two different frequencies. The lower frequencies (central ring) are connected to the brighter and distorted regions whereas the higher frequencies are characteristic of the hexagonal structure observed in graphenic regions. The inverse Fourier transform image considering only the higher frequencies (Figure 2(c)) clearly shows groups of broken bonds on the bright spot regions of Figure 2(a). Various types of dI/dV spectra were measured on the as-purchased flakes demonstrating that the band structure can be locally different on the same flake due to unsystematic distribution of atom clusters (Figure 2(a)). All the dI/dV curves shown are averaged over many dI/dV and normalized to I/V. The red dI/dV curve of Figure 2(d) depicts a wide parabolic shape with a bandgap of 200mV characteristic of dI/dV measured on oxygen group. The black dI/dV curve shows a shoulder at 0.2eV related to dI/dV carried out on close-lying oxygen surrounding. [33] For comparison the dI/dV measured at room temperature on freshly cleaved HOPG is shown in figure S3.The XPS spectrum presented in Figure 2(e) can be deconvoluted into three main peaks corresponding to C-C (284.78 eV, ~47.08%), C-O (287.03 eV, 46.19%) and C=O (288.67eV, 6.73%) bonds. Thus, the overall oxygen concentration in as-purchased GO is 52.92%.

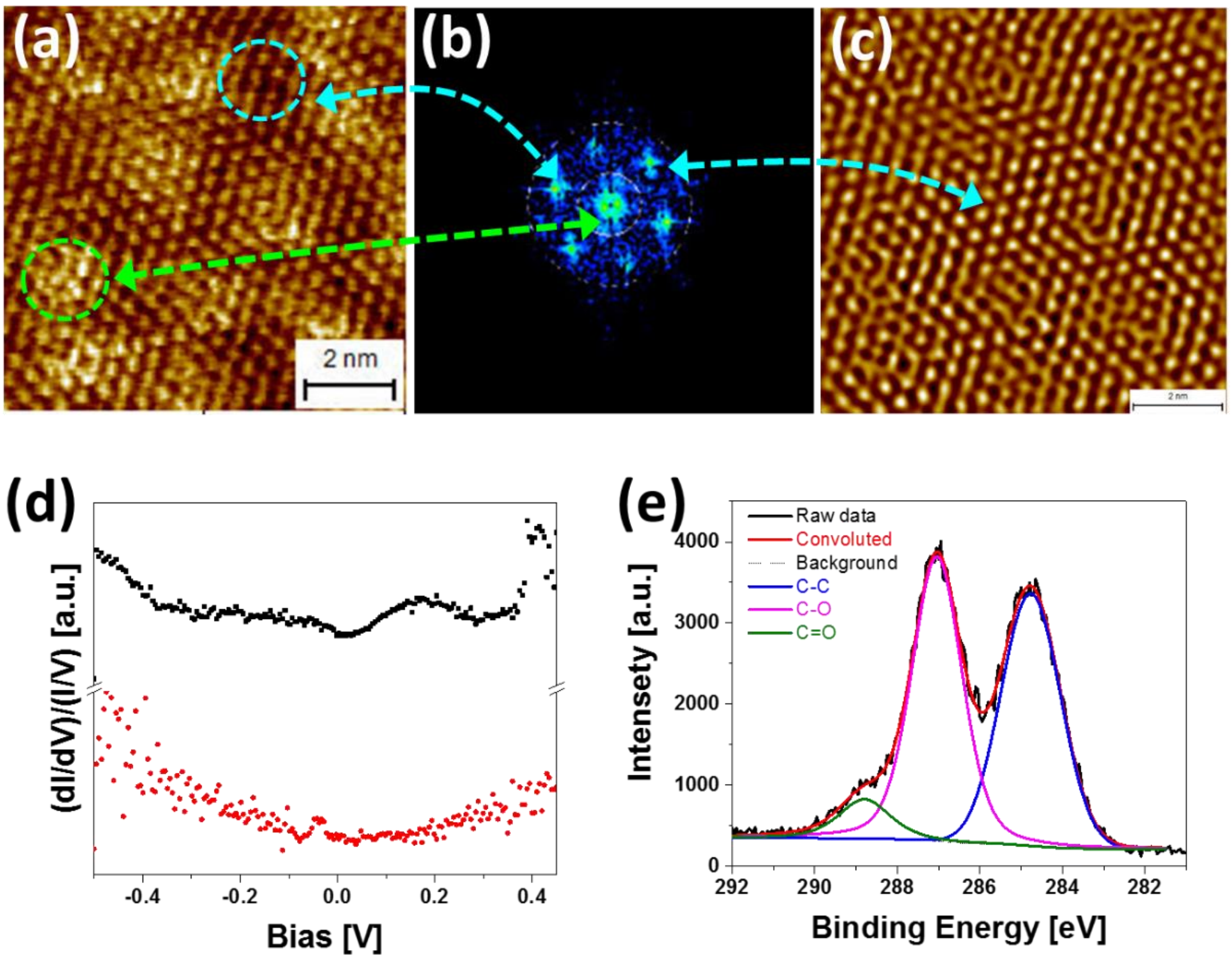

*Figure2 : Characterization of the as-purchased GO multilayer samples: (a) A STM image; (b) Fourier transform of image in (a), showing two distinct regions: An inner part at low frequency, corresponding to a distorted region as framed by a green line in (b); An outer-ring at a higher frequency, related to the "graphenic" region marked by blue-line in (a). (c) Inverse Fourier transform image considering only the higher frequencies in (b). (d) STS typical curves shows parabolic behavior (red) and shoulder at 0.2eV (black). and (e) High-Resolution Core-Level C1s XPS spectra.*

(ii) Dilution and sonication (15 minutes) of the as-purchased GO

In order to achieve GO single layers, the GO solution was first diluted and then sonicated for 15 minutes, before drop-casted onto an Au/Mica substrate. Worth noting that the dilution enabled avoiding agglomeration, however did not altered the oxygen concentration. Different sonication durations were applied, and the products were evaluated by examining the observed topography of the flakes in the AFM images. The results revealed an optimal sonication time of 15 minutes, whereas an extended procedure (~ 1 hour) led to a breakage of the flakes into tiny fragments (see SI, **Figure S2**). **Figure 3** shows the HR-TEM topographic image in real space (a), a zoom-in of the image (b) and its Fourier transform counterpart (c), of a single GO layer suspended on a TEM grid, after the 15 minutes sonication treatment of a diluted solution.

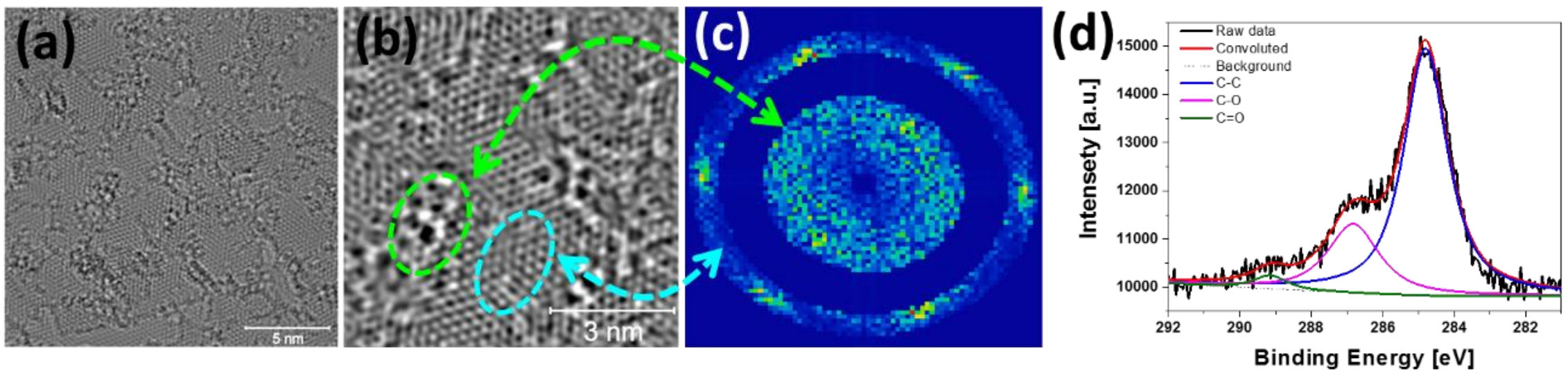

*Figure 3: (a) HR-TEM image of single GO flake; (b) A zoom-in of the image in (a); (c) Fourier transform of image in (b), showing two distinct regions: An inner part at low frequency, corresponding to a distorted region as framed by a green line in (b); An outer-ring at a higher frequency, related to the "graphenic" region marked by blue-line in (b). (d) C1s XPS spectrum of graphene oxide after 15min sonication.*

Two different frequencies can be observed on the Fourier transform. The high frequency (outer ring) corresponding to a graphenic region, as marked by the blue-circle in Figure 3(b). The lattice atomic distance extracted from the hexagonal Fourier pattern is 0.146 nm. The low frequency (central region) is associated with distorted regions on a flake, as shown by the green-circle in Figure 3(b). It should be noted that the graphenic regions appear much sharper than the distorted ones. It is most probable that the oxygen debris and oxygen functional groups concentrate on those distorted regions. The XPS spectrum (Figure 3(d)) measured after 15 minutes sonication treatment, exposed a decrease of the ratio between the C-O (286.58eV20.45%) and C-C (284.81eV, 70.60%) peaks' intensity with respect to the un-sonicated (Figure 2(b)) samples, however, the ratio of C=O (288.65eV, 8.95%) to C-C remained the same. The XPS observations are summarized in **Table 1**, reflecting an overall decrease of the oxygen concentration, originating from the C-O group. Actually, it is not clear at this moment whether the C-O group is directed bonded to a flake' surface, or clustered as a fragment on to the debris. Indeed, the HR-TEM (Figure 3(a)) exhibit pronounced ODs after the sonication stage, and therefore the type of the removed oxygen groups will be further elucidated below. We attempted to further decrease the concentration of oxygen debris by performing water rinsing.

(iii) Surplus treatment by water rinsing, after the sonication process

The GO flakes deposited on Au/Mica substrates were treated by water rinsing process (as given in the experimental section) which was applied after 15 minutes of sonication, were examined by AFM and Kelvin probe force microscopy (KPFM). **Figure 4(a)** shows the AFM image, which exhibits a topography of a representative sample with a few GO flakes of different sizes. **Figure 4**(b) displays KPFM contact potential map of the same surface as in Panel (a), indicating a higher resolution of thinnest flakes regions with respect to that shown in the AFM image. Thus, after identification of the thinnest layers, line profiles of those were plotted (see **Figure 4**(c)), from which layer's thickness of about 1-2 nm were evaluated, thus suggesting the existence of single- or double-layers after the combined treatment.

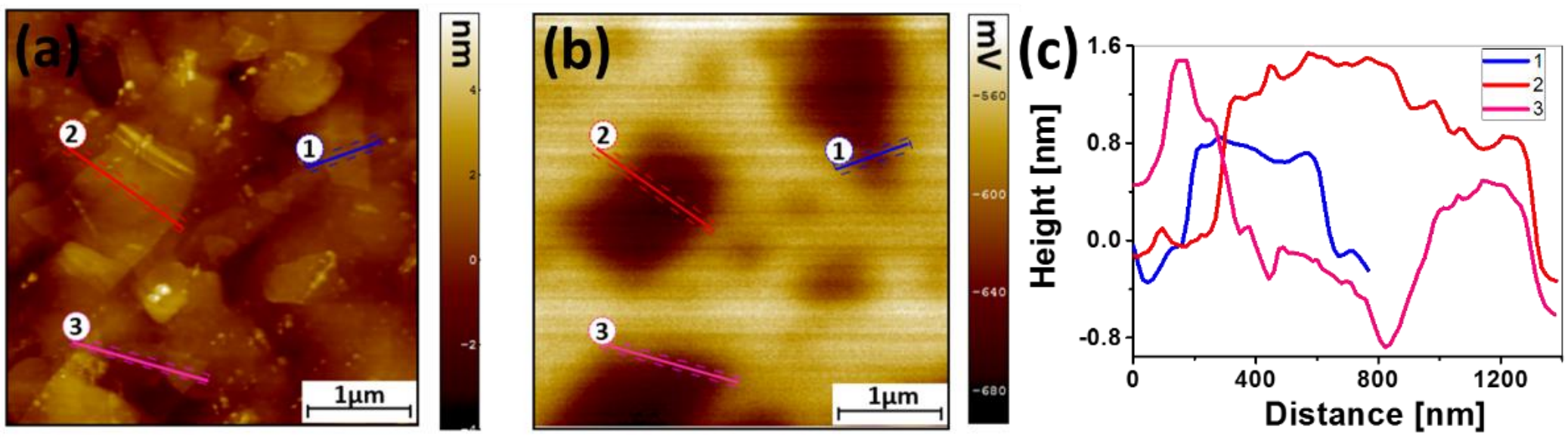

*Figure 4: (a) Atomic Force Microscopy (AFM), (b) Kelvin Probe Force Microscopy (KPFM) images, and (c) height profile of graphene oxide (GO) domains. 1, 2 and 3 are line profile of random GO domains, synchronized in images a, b and c, after sonication and rinsing (iii).*

**Figure 5** shows representative STM (a), STS curve (b) and XPS spectrum (c) of GO samples after sonication and water rinsing processes. The STM image in Figure 5(a) at a first glance resembles a pristine graphene, with substantially less distorted regions with respect to images of unrinsed samples. However, the image in Figure 5(a) shows variations in bond-lengths and appearance of some rippling after rinsing. The STS curve shown in Figure 5(b) exhibits poor fit to a parabolic shape  comperition withthat of a multilayer GO (Figure 2 (d)). The XPS spectrum (Figure 5(c)) demonstrates a weak intensity of the C-O peak (286.00eV, 4.95%) relative to that of C-C peak (284.67eV, 87.19%), thus, a smaller C-O/C-C ratio with respect that found in as-purchased and as-sonicated samples, thus revealing a further loss of oxygen atoms. In contrast, the contribution of the C=O band (287.96eV, 7.86%) remained almost unchanged upon rinsing (see Table 1).

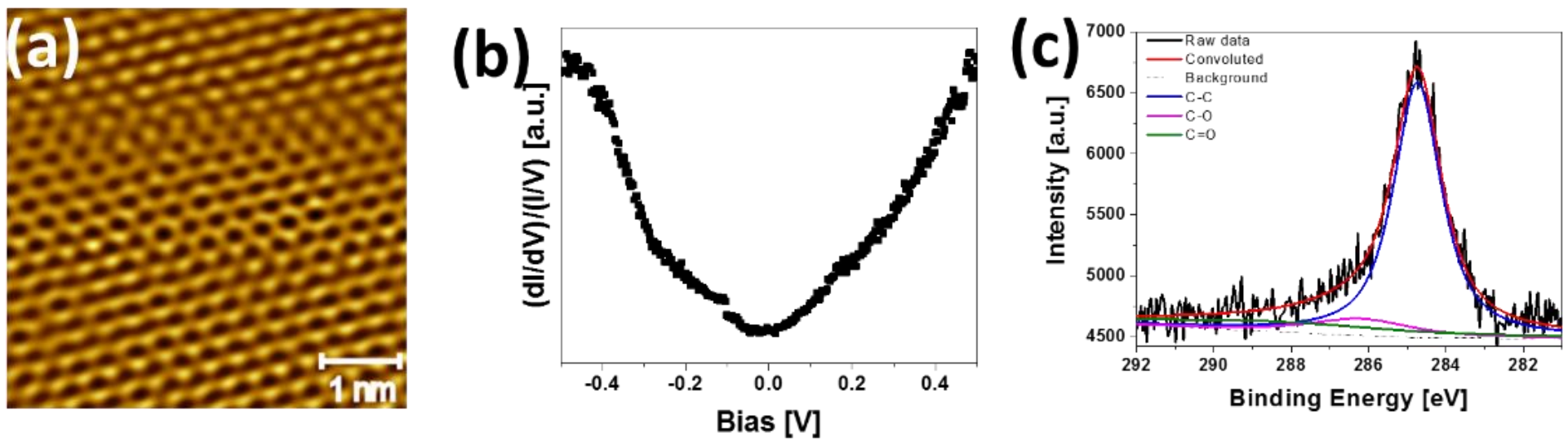

*Figure 5: (a) STM topography image of single layer and (b) STS curve from image (a). (c) High-Resolution Core-Level C1s XPS spectra of graphene oxide in a "graphenic" region, all images are after sonication and rinsing (iii).*

By analyzing the results displayed in Figure 5, several things can be noticed. The rippling seen in the STM image can be due to oxygen functional groups entrapped between the GO layer and the substrate, while the different bond lengths can be associated with oxygen functional group bonded to GO surface. The XPS results along with the diminution of the distorted regions in the STM image, suggest that part of the C-C bonds have recovered upon a removal of weakly (e.g., van der Waals) bonded oxygen groups by the water rinsing, but the covalently connected oxygen groups (e.g., C=O) stayed intact. This finding is in agreement with those published by Bonanni et al.[42] The reduction of the density of oxygen groups is further demonstrated by the typical dI/dV curve measured after the double treatment as shown in figure 5(b). It should be noted that this type of dI/dV is obtained on most of the GO flake independently of the tip location, confirming an average lower density of oxygen functional groups.

*Table 1: Relative weights of carbon species in (i) as purchased, (ii) sonicated and (iii) sonicated and water rinsed GO samples.*

| Process | Description | C-C [%] | C-O [%] | C=O [%] | C/O ratio |
|---|---|---|---|---|---|
| (i) As purchased | Multilayer GO | 47.08 | 46.19 | 6.73 | 0.89 |
| (ii) Sonicated | Single layer | 70.60 | 20.45 | 8.95 | 2.40 |
| (iii) Sonicated and water rinsed | Single layer | 87.19 | 4.95 | 7.86 | 6.81 |

Further clarification about the involved oxygen species was done by the use of Fourier transform infra-red (FTIR) spectroscopy, complementing the information gained from the XPS spectroscopy. **Figure 6** exhibits FTIR spectra of as-purchased, sonicated (15 minutes) and sonicated and water rinsed GO flakes, deposited on Au/Mica substrates, as given in the indent of the figure. For comparison, the spectra were normalized with respect to the C-C peak' intensity. The spectrum of the as-purchased GO sample is comprised of a few vibration modes, identified as the C=O, C-C, C-OH and C-O-C groups. The C-C is related to the Graphene flakes. The C=O more likely exists at the flake' rims, and other oxygen specie maybe attached to defect sites or to ODs. The FTIR spectra of the sonicated showed some reduction of the C-OH and C-O-C, with minor influence on the C=O stretching mode. More important, the surplus water rinsing after sonication, induced a substantial drop of C-OH and C-O-C peaks' intensity, along with minor effect on the strong C=O bond. The FTIR observation are compatible with the trend seen in the XPS experiments, as well as in the pronounced recovery of graphene morphology seen in the STM image (Figure 4(a)). The fact that water rinsing diminished the presence of the C-OH and C-O-C groups, suggests that these species already existed as fragments held weakly on top of OD.

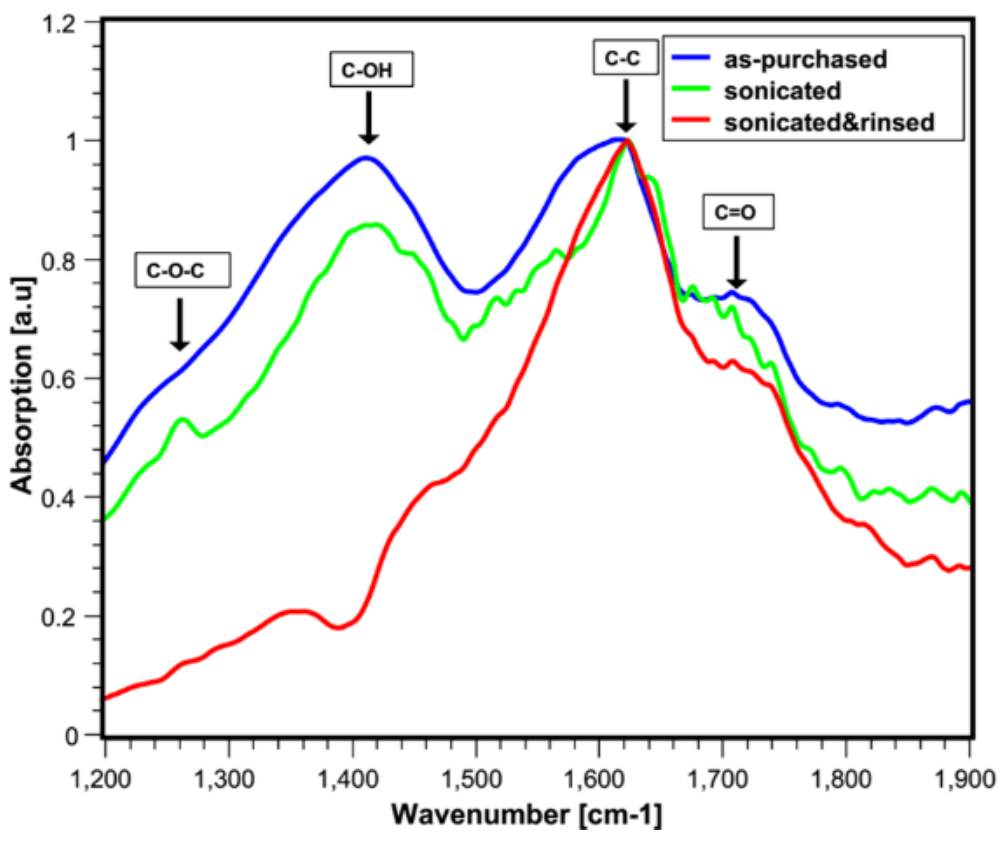

*Figure 6: FTIR spectra of GO in (i) as-purchased (blue curve), (ii) after sonication (green curve) and (iii) after sonication and water rinsing (red curve).*

Although, a substantial reduction of oxygen concentrations was shown in the XPS and FTIR measurements when progressing from as-purchased to the fully treated (sonication and water rinsing) samples, there are residues of oxygen content.  As suggested before, those residues can be related to ODs entrapped between single flakes and the Au/Mica substrate.  This assumption is further supported by the appearance of rippling, as well as by the fact that un-oxidized regions on a flake are compatible with pure graphene.  This can happen when a flake is suspended by the OD lags, avoiding direct contact with the substrate, that would normally induce strained graphene. The scheme presenting the GO flakes, which underwent evolution from the as-purchased to that found after full treatment, is shown in Figure 7.

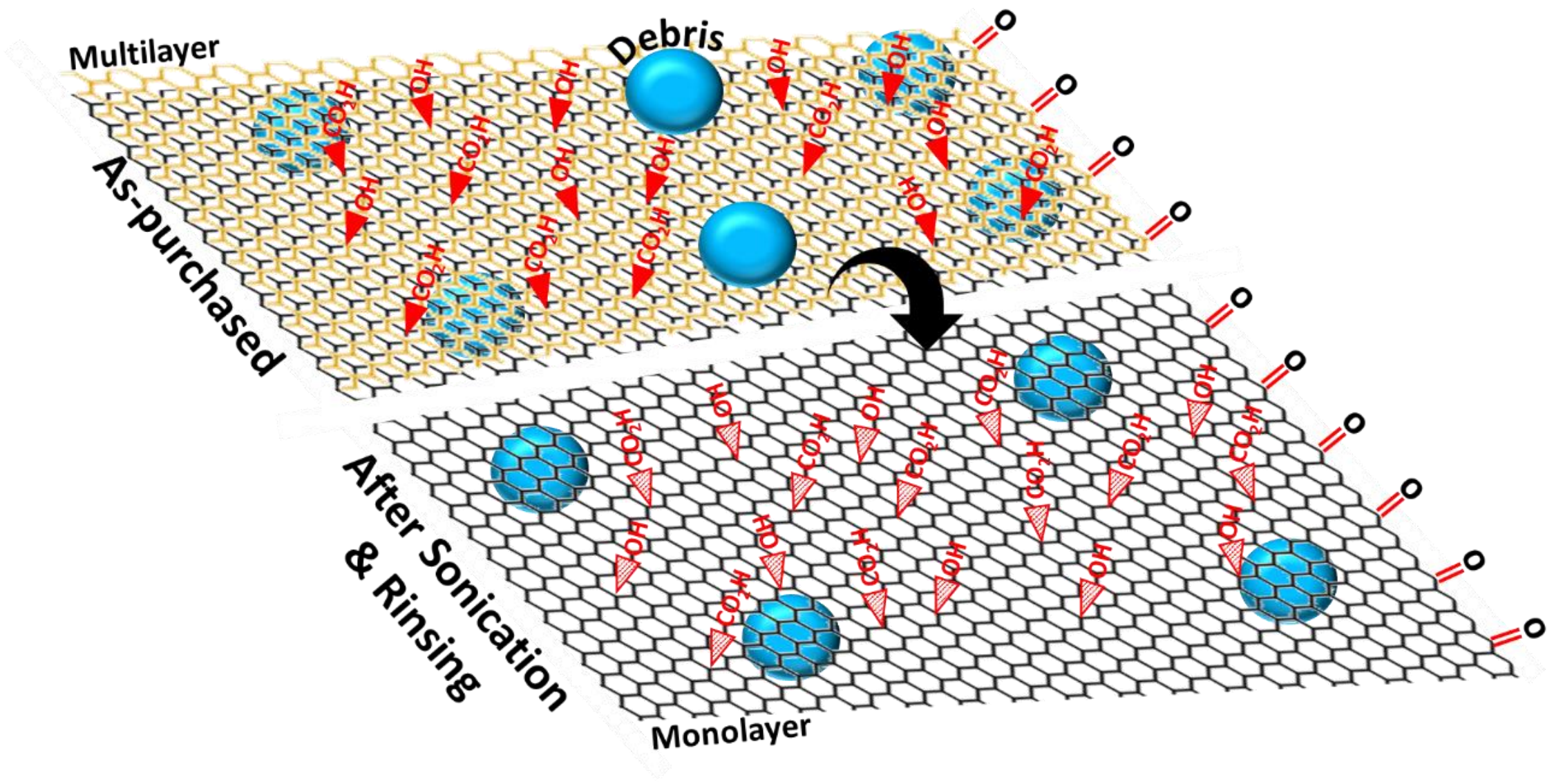

*Figure 7: The graphene oxide proposed model of as-purchased multilayer and after dilution, sonication and rinsing (monolayer). The blue balls are the oxygen clusters (also known as debris).*

## 3. Conclusion

The study analyzed the morphology and chemical composition of GO flakes which were treated by simple procedures, including short-time sonication of the as-purchased solutions, their deposition onto pre-cleaned Au/Mica substrates and finally a surplus wash with distilled water. The evolution of the flakes under the various stages were examined by microscopy (TEM, HR-TEM, STM, AFM, KPFM) and spectroscopy (STS, XPS, Raman and FTIR). The microscopy investigations of the GO flakes exhibited regions with a close resemblance to pristine graphene and at the same time, showed distinct islands composed from oxygen-related fragments (named ODs). The XPS and FTIR uncovered gradual loss of oxygen content from the as-purchased sample to that of the water-rinsed ones; in particularly the amount of C-O, C-OH and C-O-C groups had been reduced, albeit that of C=O nearly remained un-changed. The location of the C=O groups were assigned before to terminating bonds along the flakes' rims [22]. The preservation of graphene-like regions across the studied flakes occurred due to their suspension by ODs' legs, entrapped between them and the Au substrate. The STS curve bared semiconductor property beyond all treatments, owed to the oxygen atoms at the rims, the residual ODs legs or other sparse fragments. The study exposed simple and mild treatments for the regulation of GO character, which can be implemented for various graphene derivatives or other layered materials.

## Experiment

Single layer graphene oxide dispersion in ethanol, 5mg/ml, was purchased from ACS Material. The solution was diluted in ethanol absolute 99.8% H.P. and sonicated for 15min, at half frequency. Au(111)/mica substrate was purchased from PHASIS and were cleaned in ethanol and DI water. Samples of GO disposed on Au/mica substrate prepared with drop casting method. After overnight deposition, some samples were washed in distilled water to remove debris, and all samples were dried in an oven prior to microscopy and spectroscopy measurements. For Transmission Electron Microscopy (TEM) measurements, suspended GO flakes on top of ultra-thin holey carbon grids precleaned with plasma cleaner were prepared, by drop-casting diluted and sonicated GO solution.

The presence of oxygen in the films was detected by Raman and FTIR spectroscopy. The number of deposited GO layers was checked by a combination between atomic AFM and KPFM means. The distribution of the oxygen functional groups and/or oxygen debris across GO flakes were investigated by STM and HR-TEM. STM measurements were performed on a GO film deposited over a continuous Au/Mica substrate chosen for achieving low-noise STM measurements, and all the dI/dV curves shown from STS measurements are averaged over many dI/dV and normalized to I/V. HR-TEM measurements were performed on a GO film, suspended (without substrate) over holes of a TEM grid. Information about the chemical nature of the oxygen functional groups adsorbed to the surface were obtained by performing XPS and FTIR spectroscopy.

**Acknowledgements**

The authors acknowledge the financial support from the Israel Council for Higher–Focal Area Technology (No. 872967), the Volkswagen Stiftung (No.88116), the Israel Science Foundation (No.2528/19), the Israel Science Foundation (No.1045/19), USA National Science Foundation – US/Israel Binational Science Foundation (NSF-BSF, No., 2017/637). The authors would like the following colleagues for the technical assistant and for the fruitful discussions: Dr. Kamira Weinfeld from the X-ray Photoelectron spectroscopy Center, Technion, Israel; Dr. Yaron Kauffman from Electron Microscopy Center (HR-TEM), Department of Material Science and Engineering, Technion Israel; Dr. Andrey Goryachev and Prof. Yosef Raichlin from FTIR laboratory, Ariel University, Israel.